\documentclass[preprint,12pt,showkeys]{revtex4}

\usepackage[brazil, english]{babel}
\usepackage[utf8]{inputenc}
\usepackage{graphicx}
\usepackage{epsfig}
\usepackage{amssymb}
\usepackage{amsmath}
\usepackage{slashed}

\graphicspath{%
    {converted_graphics/}
    {/}
}
\begin{document}

\title{Teaching Physics in the first years of Elementary School to children with ADHD}
\author{Eduardo Folco Capossoli$^{1,}$}
\email[Eletronic address: ]{ eduardo_capossoli@cp2.g12.br}
\author{Andréa Teixeira de Siqueira Oliveira$^{2,}$}
\email[Eletronic address: ]{ andreateixeira@globo.com}
\author{Sandro Soares Fernandes$^{1,}$}
\email[Eletronic address: ]{ sandrorjbr@uol.com.br}  
\affiliation{$^1$Colégio Pedro II, Physics Teaching Department, 20.921-903, Rio de Janeiro, Brazil \\
 $^2$Elementary Education Department, 20.921-903, Rio de Janeiro, Brazil}

\begin{abstract}
In this paper we report on a Science Fair activity developed at Colegio Pedro II, a traditional Brazilian school, with a group of eight 8-12 years old Attention Deficit Hyperactivity Disorder (ADHD) students. ADHD is usually a condition associated with underachievement at school. As part of working toward scientific literacy for students, we explored the idea of conservation of energy based on STS paradigm. At the same time, the learning experience was designed to stimulate children's poor executive function, or, more specifically, their ability to manage time and planning future tasks.
\end{abstract}


\maketitle


\section{Introduction}

In Brazil, science classes in elementary school tend to emphasize Biology. Physics knowledge is hardly ever taught, even though children often show curiosity about the physical world and modern technology advances \cite{rosa}.
At a science fair, funded by CNPq (Brazilian funding agency), in 2014, at one of the most traditional schools in Rio de Janeiro, Colégio Pedro II, we had the opportunity to discuss electricity with a group of 8 -12 year-old students with learning difficulty. Once a week these students had extra classes at a Learning Laboratory in order to help them to overcome obstacles and achieve academic success. At this laboratory, students were interested in understanding how battery toys were operated, so they researched about electricity, built half a dozen electric circuits and tested the conductivity of a variety of materials. Their project, inspired by Brazilian folk culture, turned out to be a human sized rag doll with an electrical circuit quiz board on its belly. 
During those special classes, based on STS – Science, Technology and Society – paradigm, we choose to promote scientific literacy through the concept of conservation of energy. As most of the kids were diagnosed with Attention Deficit Hyperactivity Disorder (ADHD), a condition associated with poor self-regulation, they had some problem in planning what lies ahead in time. So, organizing, defining and prioritizing science fair tasks was a great opportunity to stimulate student’s executive function, or, in other words, their ability to “manage themselves effectively in order to sustain their actions (and problem-solving) toward their goals and the future” \cite{bar}.

In the reference \cite{hau} was suggested that hands-on projects can be evaluated by having the students explaining the process they experienced and why it worked the way it did. During the science fair exhibition students exposed clearly and confidently to the general public how the electric circuit quiz board worked, making it possible to evaluate as positive the impact of their learning experience. 
By taking into consideration children’s interests and their specific needs, school can become a more inclusive environment and a place where students can learn physics from its first steps.
Our intention is to share this satisfying experience with the community around the world and try to encourage other initiatives by other fellow.

\section{Colégio Pedro II}

Colégio Pedro II is a traditional school located at the city of Rio de Janeiro, Brazil. This school was founded in December 2, 1837 and belongs to the Federal Education System. Over the years, Colégio Pedro II has always been in the vanguard of education among all high schools, public and private, in Brazil.
Since September 2012, with the sanction of Law 12667/12, Colégio Pedro II is also now considered an institution of higher education. So nowadays it offers from elementary School to graduate school distributed in14 campi with around 13000 students and 1000 teachers (350 master degree and 120 doctor degree).
The present work was result of partnership between two campi, namely campus São Cristóvão I e campus São Cristóvão III.

\section{Learning Laboratory (LL)}

Special Education programs and services in Brazil are provided to students with intellectual, physical or sensory disability, autistic spectrum disorders and intellectual giftedness. Those students should preferably attend to ordinary schools and have after hours classes at Resource Rooms, once or twice a week, to learn adaptive technologies and to enhance or supplement learning. 
At Colégio Pedro II there is a Special Education Service (NAPNE) that also helps elementary school children with learning difficulty to overcome obstacles and achieve academic success. Once a week, students with learning disorders, ADHD and other issues related to underachievement in school (such as anxiety) are offered extra classes at a Learning Laboratory (LL). 
As is described in the Diagnostic and Statistical Manual of Mental Disorders, Fifth Edition (DSM-5) \cite{on}, learning disorders encompass “shortcomings in general academic skills”, including mathematics, reading and written expression (eg. Dyslexia, Dyscalculia). At LL a therapist and a teacher offer to those kids a collection of activities – computer and board games, plays, painting – rich in learning potential.
ADHD is characterized by a pattern of behavior that can result in performance issues such as “failure to pay close attention to details, difficulty organizing tasks and activities, excessive talking, fidgeting, or an inability to remain seated in appropriate situations” \cite{on2}. 
ADHD as argued in ref. \cite{bar} isn’t simply an attention deficit: it is, in fact, an executive function disorder, affecting “self-directed actions that people use to manage themselves effectively in order to sustain their actions (and problem-solving) toward their goals and the future” \cite{bar}. As it is not possible to correct directly an executive function deficit, the Learning Laboratory tries to teach ADHD students how to manage externalized forms of information and time control, in order to help them compensate for their difficulty in getting to their tasks and goals. 
One of the most successful experiences developed at the Learning Laboratory was the 2014 Science Fair Project. It was a great opportunity to stimulate ADHD students’ executive function: as most of them have trouble in planning what lies ahead in time, organizing, defining and prioritizing science fair tasks was emphasized. But, mostly, it was a great opportunity to talk about Energy and promote Scientific Literacy.
Scientific Literacy can characterize, according to \cite{lin}, two different outcomes of school science programs: developing a potential scientist (emphasis on the products, processes and characteristics of science) or “having students comprehend and cope with a variety of science- related situations” \cite{lin}. 
This latter perspective is more compatible with the STS paradigm that guided our classes: teaching science “embedded in a social and technological milieu that has scope and force for students' worlds, worldviews, or practical experiences”, helping them to understand their “natural, technological, and social worlds” \cite{jeu}.
Our emphasis at LL wasn’t on creating and testing hypotheses about electricity, but rather on playing with batteries, lamps and wires to illustrate the idea that energy can be stored on a battery and can be changed from one form (chemical) to another (electrical or luminous). These ideas were the starting point of a debate about sustainability and ways of saving energy.

\section{Preparing for the Science Fair – overview}

The Science Fair project was developed with a group of eight ADHD, 8-12 years old students. There were no home assignments and everything was carried on at LL and at the Physics Lab. Most of LL students come from economically underprivileged homes, with very little parenting support to perform academic assignments. 
The first activity was a debate to choose which experiment was going to be developed. Students were interested in understanding how battery toys were operated and they wanted to build a moving robot. A few of them also showed interest in building a potato battery just likes one they had seen on a popular tv program. 
Because there were only six weeks to prepare for the Science fair and a robot would require engineering resources that couldn’t be mobilized in such a short time, they decided to make only potato (and also lemon) batteries. They researched online how to build potato and lemon batteries, took notes, made a list of materials to be purchased. 
One week later, the experiment was all set, but the batteries didn’t work: using a multimeter they could measure a voltage as high as 2,7 volts on the potato circuit but, even though, the 1.5 volts lamp wouldn’t light up. So, that experiment was left behind (as a time management decision) and they started building a battery quiz board they had seen on a book. 
Latter, they decided to attach the electrical quiz board to a rag doll and call it a “fabric robot”. On the previous month they had made rag dolls at LL to a Literature exhibition, so they were familiarized with pattern making and the sewing process. 
To understand how the quiz board worked, they did research on books and online. At the Physics Lab they built open circuits to test materials (coins, rubber, graphite) that would close the circuit, allowing the lamp to light up. They also watched a video that showed a mechanical analogy to an electrical circuit comparing charges and basketballs. 
The students registered on a journal all learning experience and made banners to systematize a few important ideas about energy. 
On the Science Fair presentation they invited the general public – mostly High School students – to light up lamps using wires and batteries and to try to close an open circuit using materials such as coins, plastic beans and wires. The star of their presentation, though, was the fabric robot: people were invited to answer the quiz (about mathematics) and also to explain why the hidden circuit worked the way it did. 
In ref.\cite{hau} was suggested that hands-on projects can be evaluated by having the students explaining the process they experienced. As during the science fair exhibition LL students presented their experiments to the public, it was possible to observe that their knowledge about storing and changing energy from one form to another was quite solid.
Debates about sustainability and a wise use of energy took place on every class on the six weeks preparation for the Science Fair. Although it is not possible to evaluate if those ideas were assimilated to kids everyday lives, we could see they arguing about the relevance of a responsible use of energy resources at home and at school.

\section{The Science Fair}

The science fair was a result of the project submitted to CNPq, announcement MCTI/CNPq/SECIS N º 90/2013 - Science Dissemination and Popularization. The Fair took place on National Science and Technology week (SNCT), from 10/13/2015 until 10/18/2015, which belongs to the official calendar of the federal government. The science fair embraced the whole school community and family members, and had about 1500 visitors on this day.
Science fairs became popular in Brazil in the 1960s and since then the works presented in it are usually divided into three categories \cite{prog}:

\begin{itemize}
\item Assembly Works – consist the production or description of adapted mounts, usually obtained in textbooks, publications, internet etc. 

\item Informative Works – usually in the service of disseminating knowledge deemed of importance to the school community.

\item Investigative Work – associated to research projects within the many different areas of knowledge.
\end{itemize}
 
Beyond these definitions, we believe that science fairs can be understood in a similar way as in ref.\cite{man}. That is, Science fairs are social, scientific and cultural events held in schools or in the community with the intention of, during the presentation of the students, creating an opportunity for a dialogue with visitors and exhibit their creativity.

\section{Conclusions}

The worldwide prevalence of ADHD was estimated at between 5.29 \% and 7.1\%, according to DSM IV criteria \cite{pol, will}. Recent studies showed that the DSM V criteria (released on 2013) led to an increase in the prevalence rate of ADHD in US children and teenagers from 7.38\% (DSM-IV) to 10.84\% (DSM-5) \cite{van}. So, based on these references, it is possible to argue that ADHD is a reality in the world. 
In order to be inclusive and promote equity, an educational system can’t ignore this reality. ADHD doesn’t have to result in academic failure – it can be managed with appropriate educational strategies. The Science Fair Project was an effort to achieve many goals: stimulate students’ scientific curiosity, start teaching them a few physics ideas, promoting a debate about science and technology and stimulating their executive function. 
Finally, we hope to share our successful experience with other teachers who are striving to make school a more welcoming environment, a place where diversity is respected and valued.

\begin{acknowledgments}
The authors would like to thank the Brazilian agencies CAPES and CNPq for the financial support. 
\end{acknowledgments}

\end{document}